\documentclass[aps,prl,superscriptaddress,preprintnumbers,twocolumn,groupedaddress,showpacs]{revtex4}

\usepackage{graphicx}
\usepackage{hyperref}
\usepackage{amsmath}

\newcommand{\rT}{{\mathrm{T}}}

\def\mathswitchr#1{\relax\ifmmode{\mathrm{#1}}\else$\mathrm{#1}$\fi}
\newcommand{\PZ}{\mathswitchr Z}
\newcommand{\PH}{\mathswitchr H}

\newcommand{\Pe}{\mathswitchr e}

\newcommand{\Pp}{\mathswitchr p}
\newcommand{\Pj}{\mathswitchr j}

\newcommand{\Pt}{\mathswitchr t}

\newcommand{\PW}{\mathswitchr W}
\newcommand{\Pq}{\mathswitch{q}}

\def\mathswitch#1{\relax\ifmmode#1\else$#1$\fi}

\newcommand{\MW}{\mathswitch {M_\PW}}

\newcommand{\MZ}{\mathswitch {M_\PZ}}
\newcommand{\MH}{\mathswitch {M_\PH}}

\newcommand{\Mt}{\mathswitch {m_\Pt}}
\newcommand{\GW}{\mathswitch {\Gamma_\PW}}
\newcommand{\GZ}{\mathswitch {\Gamma_\PZ}}
\newcommand{\GH}{\mathswitch {\Gamma_\PH}}
\newcommand{\Gt}{\mathswitch {\Gamma_\Pt}}

\newcommand{\TeV}{\unskip\,\mathrm{TeV}}
\newcommand{\GeV}{\unskip\,\mathrm{GeV}}

\newcommand{\fb}{\unskip\,\mathrm{fb}}

\usepackage{color}

\def\draftdate{\relax}
\def\mda{\relax}
\def\mua{\relax}
\def\mla{\relax}
\def\draft{
\def\thtystars{******************************}
\def\sixtystars{\thtystars\thtystars}
\typeout{}
\typeout{\sixtystars**}
\typeout{* Draft mode!
         For final version remove \protect\draft\space in source file *}
\typeout{\sixtystars**}
\typeout{}
\def\draftdate{\today}
\def\mua{\marginpar[\boldmath\hfill$\uparrow$]%
                   {\boldmath$\uparrow$\hfill}%
                    \typeout{marginpar: $\uparrow$}\ignorespaces}
\def\mda{\marginpar[\boldmath\hfill$\downarrow$]%
                   {\boldmath$\downarrow$\hfill}%
                    \typeout{marginpar: $\downarrow$}\ignorespaces}
\def\mla{\marginpar[\boldmath\hfill$\rightarrow$]%
                   {\boldmath$\leftarrow$\hfill}%
                    \typeout{marginpar: $\leftrightarrow$}\ignorespaces}
\def\Mua{\marginpar[\boldmath\hfill$\Uparrow$]%
                   {\boldmath$\Uparrow$\hfill}%
                    \typeout{marginpar: $\uparrow$}\ignorespaces}
\def\Mda{\marginpar[\boldmath\hfill$\Downarrow$]%
                   {\boldmath$\Downarrow$\hfill}%
                    \typeout{marginpar: $\downarrow$}\ignorespaces}
\def\Mla{\marginpar[\boldmath\hfill$\Rightarrow$]%
                   {\boldmath$\Leftarrow $\hfill}%
                    \typeout{marginpar: $\leftrightarrow$}\ignorespaces}
\overfullrule 5pt
\oddsidemargin -10mm
\marginparwidth 15mm
}

\begin{document}

\preprint{Cavendish-HEP-19/15, VBSCAN-PUB-08-19}
\title{\boldmath{Exploring the scattering of vector bosons at LHCb}}
 
\author{Mathieu~Pellen}
\affiliation{Cavendish Laboratory, University of Cambridge, 
Cambridge CB3 0HE, United Kingdom}

\date{\today}

\begin{abstract}
  
In this letter, I propose a strategy to measure vector-boson scattering (VBS) at the LHCb experiment.
The typical VBS topology at hadron colliders features two energetic back-to-back jets with large rapidities and two gauge bosons produced centrally.
In this article, I show that such a topology can actually be probed by the LHCb detector.
In particular, tagging only one of the two jets in combination with two same-sign leptons allows for a measurement with upcoming luminosities.
I present an illustrative event selection where cross sections and differential distributions are computed for VBS and its irreducible background.
  
\end{abstract}


\maketitle

\subsection{Introduction}

The electroweak sector is a fascinating part of the Standard Model (SM) of particle physics.
It is imprinted by the underlying symmetries governing the SM and, in particular, the electroweak symmetry-breaking mechanism, making it a possible access point to new physics mechanisms.
One of the most exciting processes to study this is vector-boson scattering (VBS).
Due to the presence of triple and quartic gauge couplings as well as unitary cancellation, it constitutes a perfect candidate for witnessing deviations from SM expectations \cite{Contino:2010mh,BuarqueFranzosi:2017prc,Henning:2018kys}.

It is therefore paramount to measure VBS as precisely as possible and in all possible manners.
The present letter follows the latter path by devising a strategy to measure VBS at the LHCb experiment.
To my knowledge, this idea has never been promoted before and is thus completely original.
It therefore opens new opportunities for exploring scattering processes at hadron colliders and a challenging physics programme for the LHCb Collaboration.
In particular, the possibility of measuring leptons at high rapidity would allow LHCb to probe kinematic configurations that have never been explored at other experiments.
Such a measurement would thus be unique and completely independent of previous VBS measurements at the LHC.
It is thus very complementary to the measurements performed at ATLAS and CMS \cite{Aaboud:2019nmv,Sirunyan:2017ret,Aaboud:2018ddq,Sirunyan:2019der,Sirunyan:2017fvv}.

\begin{figure}
\includegraphics[width=.35\textwidth]{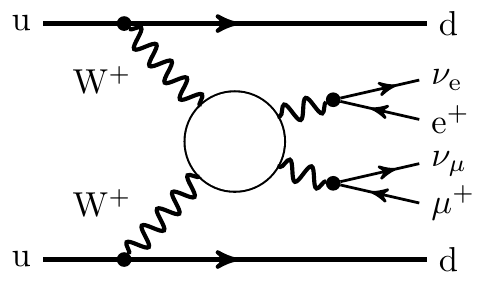}
\vspace*{-1em}
\caption{Schematic Feynman diagram representing the scattering of vector bosons at hadron colliders.
The white blob represents the VBS subprocess with {\it e.g.}\ Higgs-boson exchanges.}
\label{fig:VBS}
\end{figure}

At hadron colliders, the vector bosons scatter after being radiated off two quark lines.
A schematic Feynman diagram contributing to the process is shown in Fig.~\ref{fig:VBS}.
This particular color structure leads to a very particular topology \cite{Rauch:2016pai} where the two jets are preferably produced back to back with a large rapidity separation while the gauge bosons are produced centrally.
This feature is exploited by the ATLAS and CMS Collaborations for their measurements \cite{Aaboud:2019nmv,Sirunyan:2017ret,Aaboud:2018ddq,Sirunyan:2019der,Sirunyan:2017fvv}.
In particular, the invariant mass and the rapidity separation between the two tagging jets provide good leverage to distinguish it from its irreducible background.
At the ATLAS and CMS experiments, the golden channel is same-sign W scattering due to its large cross section in combination with a very low irreducible background \cite{Aaboud:2019nmv,Sirunyan:2017ret}.
It is followed by the WZ \cite{Aaboud:2018ddq,Sirunyan:2019der} and ZZ channels \cite{Sirunyan:2017fvv} which have lower cross sections and signal-to-background ratios but better reconstruction power.

The main challenge at LHCb is the asymmetry of the detector and thus the impossibility to reconstruct the full event unless the full system is boosted.
In addition, the luminosity at LHCb is greatly reduced with respect to the ones delivered to the ATLAS and CMS experiments.
Despite these challenges, I show in this letter that it is actually possible to measure VBS at the LHCb experiment in its future operations.
Here, I focus on the signature with one jet and two anti-muons as a prime example for the measurement\footnote{In the calculation presented below, leptons are chosen to be massless.
Therefore, a computation with electrons would be strictly identical.}.

In the first part, I motivate the event selection and strategy proposed.
I then briefly list the input parameters used for the predictions as well as the tools used.
In the third part, the cross sections and differential distributions are presented and discussed.
To conclude, I expose the main findings of this letter and ways to go beyond.

\subsection{Measurement strategy}

Typical VBS measurements rely on the fact that all final state particles are measured (the neutrinos through the missing transverse momentum).
At the LHCb experiment, the detector is only covering one part of the phase space and is asymmetric.
This implies that either the whole system has to be boosted in order to be detected as for the W+jet and Z+jet measurements \cite{AbellanBeteta:2016ugk} or only parts of the full process are detected.
In this letter, the latter avenue is followed.

\begin{figure}
\includegraphics[width=.35\textwidth]{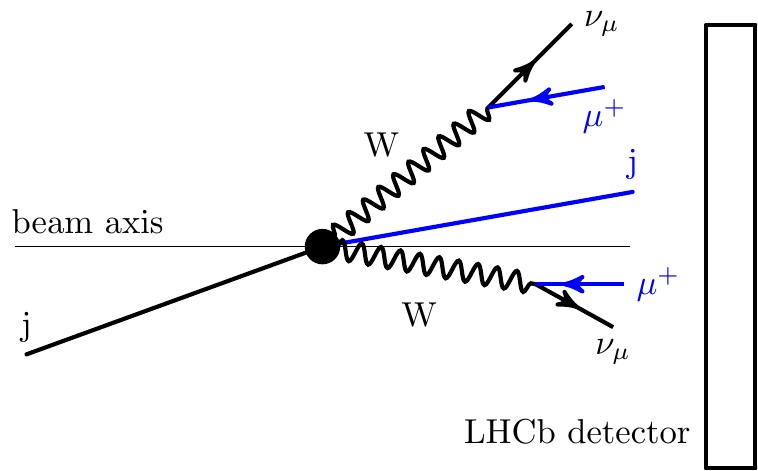}
\vspace*{-1em}
\caption{Schematic representation of a typical VBS event to be measured at the LHCb experiment.
The blue objects are the ones that are actually detected.}
\label{fig:LHCb}
\end{figure}

As mentioned previously, the golden channel for the measurement of the electroweak (EW) component of order $\mathcal{O}{\left(\alpha^{6}\right)}$ is the same-sign W channel (with the $\ell^\pm \nu_\ell \ell'^\pm \nu_{\ell'} \Pj \Pj$ final state) due to its unique signature in the SM.
Its irreducible background of order $\mathcal{O}{\left(\alpha^{4} \alpha_{\rm s}^2 \right)}$ is rather suppressed, at the level of $10\%$ while its interference of order $\mathcal{O}{\left(\alpha^{5} \alpha_{\rm s} \right)}$ 
is at the per-cent level \cite{Biedermann:2017bss}.
Therefore, it is natural to focus on measuring two same-sign leptons while tagging only one of the quark-jets \footnote{In principle, different sign could also be considered but it has the huge disadvantage of also encompassing the production of top-antitop pairs whose cross section is several orders of magnitude larger.}.
Figure~\ref{fig:LHCb} represents how such an event would be measured at the LHCb experiment.
The leptonic system is slightly boosted in order to measure the two same-sign leptons along with one of the two jets.
The second jet is not tagged as it is likely to be on the other side of the detector due to kinematic constraints.

At ATLAS or CMS, one can unambiguously distinguish between the same-sign WW (ss WW), WZ, and ZZ channels as all the final-state particles are measured.
On the other hand at LHCb, requiring same-sign leptons is not sufficient to isolate the same-sign WW and all other leptonic channels have to be included.
Indeed, one (for WZ) or two (for ZZ) charged leptons could be undetected and still lead to the signature $\ell^\pm \ell'^\pm \Pj$.

Including the WZ and ZZ channels has the drawback of lowering the signal-to-background ratio with respect to same-sign WW.
In order to diminish the effect of such channels, a veto whenever additional leptons are detected can be introduced.
From a theoretical point of view, it also has the advantage to cut away singular contributions of the type $\gamma^* \to \ell^+ \ell^-$ with low virtuality for the photon.

In principle, the final state $\ell^\pm \ell'^\pm \Pj$ with all flavour combinations $\ell, \ell' = \mu, \Pe$ should be considered.
As the present study is mainly illustrative, only the case $\mu^+ \mu^+$ is examined here.
It is justified by the fact that the cross section of the negative signature is only about one third of the positive one due to different parton distribution functions (PDF) \cite{Chiesa:2019ulk}.

To be more concrete, the event selection reads as follows.
The final state is $\mu^+ \mu^+ \Pj$ and the requirements on these objects are:

\begin{eqnarray}
\label{eq:cut}
 p_{\rT,\rm j} &{}>& 20\GeV, \quad 2.2 < \eta_{\rm j} < 4.2,\\
 p_{\rT,{\mu^+}} &{}>& 20\GeV, \quad 2.0 < y_{\mu^+} < 4.5, \\
 \Delta R_{\Pj\mu^+} &{}>& 0.5.
\end{eqnarray}

Thanks to the high-rapidity coverage of the LHCb detector for leptons, it can reach kinematic regions that are not accessible to ATLAS or CMS.
These regions receive very large EW corrections \cite{Biedermann:2017bss} which are thus interesting to explore in the SM and beyond.
In addition to the above cuts, a veto is applied to all events featuring extra lepton(s) of different charge or flavour in the detector region
\begin{eqnarray}
\label{eq:veto}
2.0 < \eta_{\rm \ell} < 4.5,
\end{eqnarray}
with $\ell = \mu^-, \Pe^+, \Pe^-$.
Its main purpose is to reject as much as possible the $\PW\PZ$ and $\PZ\PZ$ contributions which have worse signal-to-background ratios than ss-$\PW\PW$.

The philosophy of this event selection is to suppress as much as possible contributions other than the ss-WW one.
This implies that the signal over background is maximal but the overall statistics are lower.
Relaxing these requirements could improve the measurement and therefore the present selection should be understood as a pessimistic scenario.
For example, both muons have been required to have a transverse momentum larger than $20\GeV$ for simplicity while in reality the trigger requires only one of them.
But such optimisations of the cuts require experimental knowledge on efficiencies, event yields, fake backgrounds and therefore go beyond this exploratory theoretical work.
These will be addressed in a dedicated study \cite{Kenzie}.

\subsection{Details of the calculation}

Given that all channels contribute to the final state $\mu^+ \mu^+ \Pj$, the following hadronic processes have been simulated:
\begin{eqnarray}
\label{eq:process}
\Pp \Pp &\to& \mu^+ \nu_\mu \mu^+ \nu_{\mu} \Pj \Pj \quad ({\rm ss \hspace{0.1cm} WW}), \\
\Pp \Pp &\to& \mu^+ \nu_\mu \mu^+ \mu^- \Pj \Pj \quad ({\rm WZ}), \\
\Pp \Pp &\to& \mu^+ \mu^- \mu^+ \mu^- \Pj \Pj \quad ({\rm ZZ}),
\end{eqnarray}
at orders $\mathcal{O}{\left(\alpha^{6}\right)}$ (denoted by EW).
These are the signal processes containing VBS contributions.
The dominant irreducible QCD backgrounds (denoted by QCD) for these processes are:
\begin{eqnarray}
\Pp \Pp &\to& \mu^+ \nu_\mu \mu^+ \nu_{\mu} \Pj \Pj, \\
\Pp \Pp &\to& \mu^+ \nu_\mu \mu^+ \mu^- \Pj, \\
\Pp \Pp &\to& \mu^+ \mu^- \mu^+ \mu^- \Pj,
\label{eq:bckprocess}
\end{eqnarray}
at orders $\mathcal{O}{\left(\alpha^{4} \alpha_{\rm s}^2 \right)}$ and $\mathcal{O}{\left(\alpha^{4} \alpha_{\rm s} \right)}$ (for the last two).

Note that for the EW contributions, singular contributions can also arise from $\gamma^* \to \Pq \bar \Pq$ subprocesses in the WZ and ZZ channels.
In the simulations, these have been regulated by technical cuts as their effects are small \cite{Denner:2019tmn,Denner:2019zfp}.
Nonetheless, for completeness, they should be dealt with using the method proposed in Ref.~\cite{Denner:2019zfp}.

Also, the interference contribution of order $\mathcal{O}{\left(\alpha^{5} \alpha_{\rm s} \right)}$ has been left out in this study as it usually amounts to just a few per cent \cite{Biedermann:2017bss,Denner:2019tmn}.
All predictions are made at leading order (LO).
To obtain the subleading QCD contributions at order $\mathcal{O}{\left(\alpha^{4} \alpha_{\rm s}^2 \right)}$ in the channels WZ and ZZ, the next-to-leading order (NLO) QCD corrections should be computed.
For V+j, in a similar set-up, they have been found to be about $+30\%$ \cite{Gehrmann-DeRidder:2019avi}.

For all predictions, the resonant particles are treated within the complex-mass scheme~\cite{Denner:1999gp,Denner:2005fg}, ensuring gauge invariance.
To evaluate all tree amplitudes in the 5-/6-body phase space, the computer code {\sc Recola}~\cite{Actis:2012qn,Actis:2016mpe} is employed.
The integration is performed with the Monte Carlo program {\sc MoCaNLO} which has already been used in NLO computations for VBS \cite{Biedermann:2016yds,Biedermann:2017bss,Ballestrero:2018anz,Denner:2019tmn}.

Theoretical predictions are presented for $\Pp\Pp$ collisions at a center-of-mass energy of $13\TeV$.
The on-shell values for the masses and widths of the gauge bosons read
\begin{equation}
\begin{array}[b]{rcl@{\quad}rcl}
  \MW^{\rm os} &=& 80.379  \GeV, & \GW^{\rm os} &=& 2.085 \GeV, \\
  \MZ^{\rm os} &=& 91.1876 \GeV, & \GZ^{\rm os} &=& 2.4952\GeV
\end{array}
\end{equation}
and are converted into pole masses according to
\begin{eqnarray}
&& M_V = M_V^{\rm os}/c_V, \quad \Gamma_V = \Gamma_V^{\rm os}/c_V, \nonumber\\
&& c_V=\sqrt{1+(\Gamma_V^{\rm os}/M_V^{\rm os})^2}, \quad V=\PW,\PZ.
\end{eqnarray}
The Higgs-boson and top-quark masses and widths are fixed to 
\begin{equation}
\begin{array}[b]{rcl@{\quad}rcl}
\MH &=& 125\GeV,         & \GH   &=& 4.07 \times 10^{-3}\GeV, \nonumber \\
\Mt &=& 173\GeV,         & \Gt   &=& 0\GeV .
\end{array}
\end{equation}
The top-quark width has been set to zero as no resonant top quarks appear at tree level when no external bottom quarks are considered.

For the electromagnetic coupling $\alpha$, the $G_\mu$ scheme is
used where $\alpha$ is obtained from the Fermi constant,
\begin{equation}
\alpha_{G_\mu} = \sqrt{2}G_\mu\MW^2\left(1-\MW^2/\MZ^2\right)/\pi ,
\end{equation}
with
\begin{equation}
G_\mu= 1.16638\times 10^{-5} \GeV^{-2}.
\end{equation}

The PDF set NNPDF31\_lo\_as\_0118 \cite{Ball:2017nwa} has been used everywhere \footnote{PDF uncertainties for VBS have been shown to be well under control in Refs.~\cite{Bellan:2019xpr,Schwan:2018nhl}.}.
The scale $\mu$ is set to the pole mass of the W boson, $\mu = M_W$.
Quarks and gluons are clustered using the anti-$k_\rT$ algorithm
\cite{Cacciari:2008gp} with jet-resolution parameter $R=0.4$.

\subsection{Numerical results}

First, the cross sections for the processes \eqref{eq:process}-\eqref{eq:bckprocess} in the set-up of Eqs.~\eqref{eq:cut}-\eqref{eq:veto} are given in Table~\ref{tab:xsec} in femtobarns.
In addition, the ratios $\sigma_{\rm EW}/\sigma_{\rm QCD}$ are also given.

\begin{table}
\begin{center}
\begin{tabular}
{cccc}
\hline
Channel & $\sigma_{\rm EW}$~[fb] &  $\sigma_{\rm QCD}$~[fb] & $\sigma_{\rm EW}/\sigma_{\rm QCD}$
\\
\hline
ss WW & $0.0185(1)$ & $0.0104(1)$& $1.78$ 
\\
\hline
WZ & $0.0071(1)$ & $0.2952(4)$& $0.02$ 
\\
\hline
ZZ & $0.0003(1)$ & $0.0161(1)$& $0.02$ 
\\
\hline
Sum & $0.0258(1)$ & $0.3217(4)$ & $0.08$ 
\\
\hline
\end{tabular}
\end{center}
\caption{Cross sections for processes contributing to $\Pp \Pp \to\mu^+\mu^+\Pj +X$ at $13\TeV$ at LHCb.
The cross sections are expressed in femtobarn for the orders $\mathcal{O}{\left(\alpha^{6}\right)}$ (EW) and $\mathcal{O}{\left(\alpha^{4} \alpha_{\rm s}^2 \right)}$ or $\mathcal{O}{\left(\alpha^{4} \alpha_{\rm s} \right)}$ (QCD).
The digit in parenthesis indicates the integration error.}
\label{tab:xsec}
\end{table}

As expected, for the EW component, the cross sections are larger for processes with W instead of Z couplings.
As for the ATLAS and CMS measurements, the same-sign WW channel is clearly the golden channel to measure VBS in terms of cross section and background.
Finally, the last line where the sum over all channels is performed is the physical cross section that would be measured in the experiment when looking at the $\mu^+ \mu^+ \Pj$ final state.
It amounts to about $0.35\fb$ and is the combined cross section of the EW ($8\%$) and QCD ($82\%$) contributions.
Using scale variation by a factor 2, the estimated theoretical error is $[+4.6\%, -4.3\%]$ on the EW component and $[+10.2\%, -9.1\%]$ on the QCD component.

I stress again that the results presented here correspond to a pessimistic scenario where the signal-to-background ratio is large while the statistics is limited.
Also, here only one event selection has been presented while one could devise two strategies depending on whether one wants to measure the combine process or only the EW component.
To that end, a more in-depth study should be performed based on the detailed knowledge of the LHCb detector \cite{Kenzie}.

For illustrative purposes, only the case $\mu^+\mu^+$ has been considered here.
In the limit of massless leptons, $\sigma_{\mu^+\mu^+} = \sigma_{\Pe^+\Pe^+}$.
In addition, given that interference contributions are negligible \cite{Chiesa:2019ulk},
$\sigma_{\PW\PW\to\Pe^+\mu^+} \simeq 2 \sigma_{\PW\PW\to\mu^+\mu^+}$,
$\sigma_{\PW\PZ\to\Pe^+\mu^+} \simeq   \sigma_{\PW\PZ\to\mu^+\mu^+}$, and
$\sigma_{\PZ\PZ\to\Pe^+\mu^+} \simeq 2 \sigma_{\PZ\PZ\to\mu^+\mu^+}$.
This implies that the total combined cross section (QCD+EW) is about
\begin{equation}
 \sigma_{\ell^+\ell'^+} \simeq 1.4\fb,
\end{equation}
with $\ell, \ell' = \mu, \Pe$.
From these, $7.5\%$ {\emph i.e.}\ about $0.1\fb$ is due to the EW production.
In addition, the cross section with negatively charged leptons can also be considered using the same principle.
Even if it represents only a fraction of the \emph{++} signature due to PDF contributions \cite{Chiesa:2019ulk}, it has the same diagrammatic contributions and thus is equally interesting.
With an expected luminosity of $50\fb^{-1}$ or even $300\fb^{-1}$ for future operations of LHCb, measuring both the combined QCD and EW contributions as well as the EW component on its own is promising.
To that end, a combined measurement which is tested against a hypothesis with and without an EW component is preferred over a measurement where the QCD contribution is subtracted from the data based on Monte Carlo simulations.
Indeed, as pointed out in Ref.~\cite{Biedermann:2017bss}, the notion of EW signal and QCD background is ill-defined at NLO from a theoretical point of view due to interferences.

\begin{figure}
\hspace{-2cm}
\includegraphics[width=.47\textwidth]{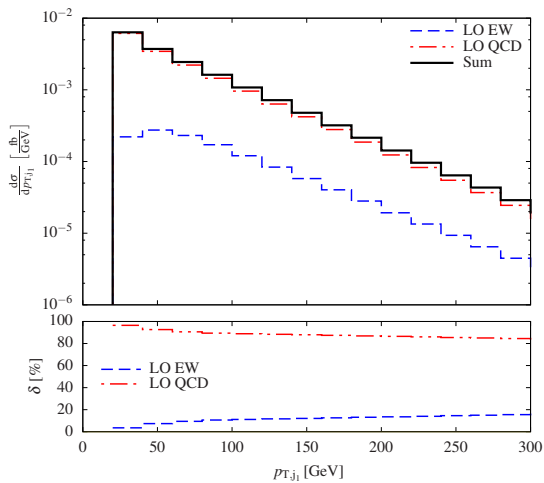}
\vspace*{-1em}
\caption{Transverse-momentum distribution of the reconstructed jet for $\Pp \Pp \to\mu^+\mu^+\Pj +X$ at $13\TeV$ at LHCb.
The QCD (red) and EW (blue) components are shown in absolute value (upper panel) and relative to their sum (lower panel).}
\label{fig:diff1}
\end{figure}

\begin{figure}
\hspace{-2cm}
\includegraphics[width=.47\textwidth]{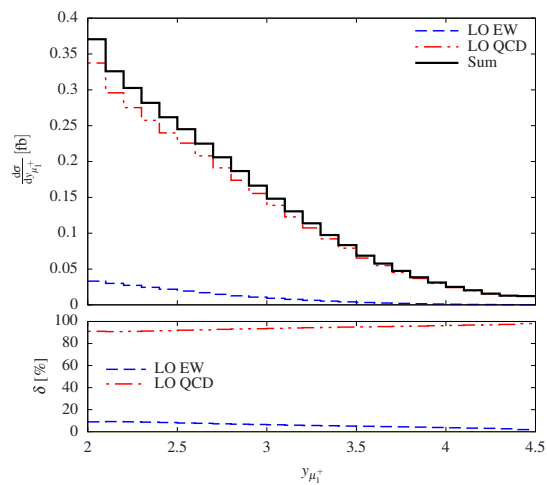}
\vspace*{-1em}
\caption{Rapidity distribution of the hardest anti-muon for $\Pp \Pp \to\mu^+\mu^+\Pj +X$ at $13\TeV$ at LHCb.
The QCD (red) and EW (blue) components are shown in absolute value (upper panel) and relative to their sum (lower panel).}
\label{fig:diff2}
\end{figure}

Finally, two differential distributions are also shown in Figs.~\ref{fig:diff1} and \ref{fig:diff2}.
In the upper plot, the absolute predictions for the EW and QCD components as well as their sum is shown for all channels together.
The lower plot shows the contributions of the EW and QCD components with respect to the combined process.

Both distributions show that the composition of the combined process is not uniform over the kinematic range displayed.
Figure~\ref{fig:diff1} shows that the EW contribution steadily increases toward high transverse momentum of the hardest jet to reach about $20\%$ at $300\GeV$.
On the other hand, for the rapidity of the hardest anti-muon (see Fig.~\ref{fig:diff2}), the maximal EW composition is reached for the minimum rapidity (here $2.0$ due to the detector limitations).
While these distributions are mainly illustrative here, they suggest ways to improve the signal-to-background ratio in certain phase-space regions.

\subsection{Discussion}

In this article, I have presented an exploratory study for the measurement of VBS processes at the LHCb experiment.
In particular, an event selection has been designed to deal with the unique design of the LHCb detector.
The key point is that not all final states are required to be tagged as opposed to what is traditionally done at ATLAS or CMS.
Based on this set-up, numerical simulations of the signal and background processes have been performed.
The results are promising and show that a combined measurement of the QCD and EW components, and even of the EW contribution on its own, can be in reach for the high-luminosity runs of LHCb.

While the present study provides the main idea and first theoretical inputs for such a measurement, it can be extended in several ways.
First, it would be desirable to have NLO QCD and EW corrections for both the signal and the background along the lines of Refs.~\cite{Biedermann:2017bss,Denner:2019tmn}.
Note that this task is rather challenging as not only QCD corrections but also EW corrections should be computed as these are large for VBS at the LHC \cite{Biedermann:2016yds}.
Including the background processes, it amounts thus to compute 12 NLO computation, some of which are still unknown.
Second, a more thorough analysis of the experimental capabilities should be performed.
It would be interesting to optimise the event selection depending on whether one wishes to target a combined measurement or a measurement of the EW component only.
In particular, one should explore the possibilities for the different flavour and charge combinations as well as provide a detailed estimation of the experimental systematic errors \cite{Kenzie}.
Finally, it would be important to investigate whether specific new-physics models are enhanced in the LHCb kinematic and could then be stress-tested with such a measurement.

The asymmetric nature of the LHCb detector and the low luminosity available constitute the main challenges to overcome.
Nonetheless, this measurement would allow one to test the SM even further and possibly explore its connections with new mechanisms.
In particular, the LHCb detector allows for measurements in the forward region which is very different from the kinematic usually probed at other experiments.
Performed in a unique environment, such an independent measurement would hence complement the existing ones very well.

\vspace{1em}

\subsection{Acknowledgements}

I would like to thank Matthew Kenzie for useful discussions on the LHCb experiment and comments on the manuscript.
This project has received funding from the European Research Council (ERC) under the European Union's Horizon 2020 Research and Innovation Programme (grant agreement No. 683211).

\bibliography{vbs_LHCb}

\end{document}